\documentclass{article}

\usepackage{arxiv}

\usepackage[utf8]{inputenc} 
\usepackage[T1]{fontenc}    
\usepackage{hyperref}       
\usepackage{url}            
\usepackage{booktabs}       
\usepackage{amsfonts}       
\usepackage{nicefrac}       
\usepackage{microtype}      
\usepackage{lipsum}
\usepackage{graphicx}
\graphicspath{ {./images/} }
\usepackage{array}

\usepackage{cite}
\usepackage{mdwmath}
\usepackage{mdwtab}

\usepackage{amssymb}
\usepackage{cite}
\usepackage{eucal}
\usepackage{graphicx, multicol, latexsym, amsmath}
\usepackage{graphics}
\usepackage{blindtext}
\usepackage{subfigure}
\usepackage{caption}
\usepackage{capt-of}
\usepackage[noend]{algpseudocode}
\usepackage[boxruled]{algorithm2e}
\usepackage{enumitem}
\usepackage{multirow}
\usepackage{tabularx}
\usepackage{adjustbox}
\usepackage{float}
\usepackage{eqparbox}
\usepackage{makecell}
\usepackage[noend]{algpseudocode}
\usepackage{multirow}
\usepackage{mathtools}
\DeclarePairedDelimiter\ceil{\lceil}{\rceil}
\DeclarePairedDelimiter\floor{\lfloor}{\rfloor}

\title{AppSign: Multi-level Approximate Computing for Real-Time Traffic Sign Recognition in Autonomous Vehicles}

\author{
 Fatemeh Omidian \\
  Faculty of Computer Engineering\\
  K. N. Toosi University of Technology\\
  Tehran, Iran. \\
   \And
 Athena Abdi \\
  Faculty of Computer Engineering\\
  K. N. Toosi University of Technology\\
  Tehran, Iran. \\
  \texttt{a\_abdi@kntu.ac.ir} 
}

\begin{document}
\maketitle
\begin{abstract}
This paper presents a multi-level approximate computing approach for real-time traffic sign recognition in autonomous vehicles called AppSign. Since autonomous vehicles are real-time systems, they must gather environmental information and process them instantaneously to respond properly. However, due to the limited resources of these systems, executing computation-intensive algorithms such as deep-learning schemes that lead to precise output is impossible and takes a long time. To tackle this, imprecise computation schemes compromise the complexity and real-time operations. In this context, AppSign presents a multi-level approximate computing scheme to balance the accuracy and computation cost of the computation-intensive schemes and make them appropriate for real-time applications. AppSign is applied to the CNN-based traffic sign recognition unit by approximating the convolution operation of CNN which is the primal solution for image processing applications. 
In AppSign a novel approximate multiplication method called "TIRuD" is proposed that truncates the operations while keeping the accuracy acceptable. Moreover, it provides the adaptive approximation of the underlying CNN by involving various levels of computation and considering different approximation methods.
The efficiency of the proposed AppSign, in real-time traffic sign recognition, is evaluated through several experiments. Based on these experiments, our proposed TIRuD reduces the accuracy by about $10\%$ while saving execution time about $64\%$ over the exact multiplication, averagely. Moreover, employing our proposed hierarchical approximation in various model layers outperforms the exact computation $27.78\%$ considering "AoC" that joins accuracy and computation cost in a parameter.      
\end{abstract}


\section{Introduction}
\label{sec:intro} 
Embedded systems are information processing units integrated into a larger system that includes various hardware and mechanical components and serves a specific purpose. Due to their proper performance, embedded systems are now utilized in various medical, commercial, military, industrial, transportation, communication, and more applications. Efficiency is one of the most important requirements of these systems that are provided with limited resources performing a dedicated task optimally~\cite{marwedel2021embedded,embeddedbook2024}. Autonomous Vehicles are a common use of embedded systems in transportation that provide various advantages and are designed with multiple levels of automation\cite{understanding, AV-review}. 

Self-driving cars are intelligent vehicles that receive environmental information through embedded hardware and software. They analyze and organize this gathered information to execute the necessary control responses in time and without human intervention. The correct and timely recognition of the environmental signs and analysis of them while driving is important to prevent accidents, congestion, and traffic. Traffic signs are one of these environment signals to recognize the environment better, adjust the actions, and control responses based on them~\cite{a1, understanding, AV-review}. Recognition of the traffic signs is easily performed by humans while driving, but in self-driving cars, accurately and quickly locating and classifying the guiding signs is a major challenge that has caught the attention of computer vision researchers. The prominent efficiency of deep learning approaches in computer vision applications such as traffic sign recognition, makes them an appropriate solution that provides high accuracy. However, these approaches apply significant processing cost and time to the underlying system which is unacceptable in resource-constraint embedded systems. In this context, embedded machine vision approaches are proposed to solve these challenges and apply accurate methods in embedded applications that require limited resources and processing time \cite{a3, TSR-2023,a5,a7}. 

Employing computation-intensive vision schemes in resource-constraint embedded systems such as autonomous vehicles, challenges the designers. Several solutions based on dynamic algorithm configuration, hardware compression, and computing compression are presented in this context. Since convolution neural networks (CNN) are one of the most effective and primal deep learning-based approaches for image processing, most research is focused on it. 
Dynamic configuration adapts the CNN layers for different inputs to reduce the computations as required~\cite{dca1,a9}. Hardware compression schemes customize the design of hardware units such as multipliers to perform efficient computations considering execution time, area, and energy consumption. These accelerators employ a high level of parallelism to consider the repetitive operations of the target unit~\cite{accel1,accel2}. Besides the algorithm and hardware compression schemes, other approaches focus on computations. Approximate computing approaches simplify the operations and make them inaccurate while having acceptable accuracy. These schemes utilize the gap of precision required by applications and that provided by the computing system to adapt the cost and overhead of the target system \cite{app-cnn,a13,appcomp1}. 

Approximate computing is an appropriate solution to regulate the computation overhead in applications where the best accuracy is not always required \cite{a11,appcomp1,app-cnn}.
Multimedia applications including image processing that are tightly involved with computationally intensive deep learning schemes are appropriate candidates for approximation. Multiplication is the common computation of these algorithms which is complicated and consumes significant energy. This makes this operation a suitable candidate for the approximate design that saves time on the cost of imprecision which is tolerable in limited scale~\cite{a11,a12}.
Various methods with different accuracy and efficiency levels have been presented that can be further divided into three families: approximate multipliers/adders, approximate multiplication operators, and approximate log multipliers~\cite{cir2024,appmul,a13}. 
These approximations regulate the accuracy with respect to computation overhead that is useful in implementing image processing algorithms in embedded applications such as self-driving cars. However, their improvement and appliance in real-time systems are still an open research field.

Our proposed AppSign method aims to develop an embedded system for traffic sign recognition in self-driving cars.
It utilizes an approximated deep learning-based scheme to balance accuracy, computational overhead, and response time. In AppSign a CNN-based image classifier architecture is considered and its convolutions are approximated, efficiently. 
This system is real-time and its response time is directly affected by the computation overhead. Moreover, the accuracy and computation overhead have an antagonistic relation. AppSign determines the approximation scope and identifies its appropriate layers, to meet this trade-off efficiently.  
In AppSign an approximate multiplier called "TIRUD" is proposed to reduce the computation overhead while providing high accuracy by truncating and rounding the operands. Moreover, the sensitivity of the operations of various layers of the deep structure on the final result is examined by applying different approximation methods. In this way, various approximation methods are combined non-uniformly to provide high accuracy while having low computation overhead.   
The main contributions of our proposed AppSign can be summarized as follows: 

\begin{itemize}
    \item Presenting an approximate sign detection unit called "AppSign" for self-driving cars based on approximate convolution operations in CNN structure;
    \item Presenting an approximate multiplier called "TIRuD" that adjusts the precision of the computation based on the importance of each part of data to reduce computation while regarding accuracy in convolution operation; 
    \item Employing a non-uniform and multi-layer approximation scheme on sign detection unit to balance the accuracy and computation overhead through combining various high and low precision approximation methods;
    \item Presenting a metric called "AoC" to consider the accuracy and computational cost/time jointly to compromise these contradicted parameters during design more efficiently. 
\end{itemize}

The rest of the paper is organized as follows: Section 2 presents the literature review, and Section 3 explains the details of the proposed AppSign considering the approximation method and sign recognition unit. Section 4 presents the experimental results and Section 5 includes the concluding remarks and future trends.  

\section{Related work}
\label{related}
Embedded systems are widely used in modern applications and are resource-constraint due to their efficiency necessity in performance and energy consumption. Moreover, the dependability of these systems is important due to their appliance in safety-critical applications. 
To this aim, deep learning schemes that provide acceptable precision when facing big data are employed. These schemes are computation-intensive while the embedded systems have limited processing resources. To tackle this, various research projects are proposed and classified into dynamic algorithm configuration, hardware compression, and computation compression.

The dynamic configuration approaches aim to simplify the algorithms by reducing unnecessary computations. The computations of CNN layers are not always utilized for simple and complicated tasks. Thus, a dynamic layered CNN structure based on the input is introduced in~\cite{dca1}.
In addition to dynamic layered CNNs, the YOLO (You Only Look Once) algorithm is another example of these methods~\cite{a15}. YOLO utilizes a neural network to predict bounding boxes and class probabilities from full images in a single assessment. Neural Architecture Search (NAS) is employed to optimize the performance of a neural network for a given task by exploring the search space for potential architectures and pinpointing those that produce superior results~\cite{a16}. In addition, neural networks can be dynamically configured using methods like conditional computing~\cite{a17} and distributed computing~\cite{a18} that selectively activate the network parts based on the input need. 

Hardware compression-based approaches aim to design specialized and high-performance hardware for certain repetitive operations to improve their efficiency in computations. 
For instance, in the multiplication operation, the multiplier and the product can be reused efficiently to reduce computation costs. A LUT-based multiplexing structure has been proposed in~\cite{acc3} to decrease energy consumption. 
Furthermore,~\cite{a19} outlines a novel method for constructing deep neural network (DNN) accelerators utilizing quantum-dot cellular automata (QCA). A hardware accelerator based on combined pruning and non-pruning layers of CNN to achieve a tradeoff between compression ratio and processing efficiency at a small loss in accuracy is presented in~\cite{a20}. 

The third category of methods aims at compressing the computation. Hardware DNN approximations can be classified into three general categories: precision scaling, computation reduction, and approximate arithmetic units \cite{a13}.
One of the first and most widely used approximation techniques to enable effective precision scaling is quantization. Quantization shortens integer operations and saves computation costs. 
These implementations linearly reduce the model size, increasing storage and decreasing memory throughput~\cite{a13,a21,a22,a23}.
Multi-level quantization is proposed to reduce the computations in two levels, first statically and then selectively considering some of the less informative layers in the network~\cite{a24}. 

Furthermore, in computation reduction methods the objective of approximation considers minimizing calculations while avoiding specific calculations, such as multiplication and convolution operations. Methods for computation reduction include skipping and memorization approximation~\cite{a13}. 
The widely utilized ReLu activation function is employed in~\cite{a33} to eliminate redundant computations. In \cite{a35}, the near-zero approximation (NZA) was introduced by counting the initial zeros of the multiplication operands, and if their number is greater than a threshold, the product will be deemed zero and the multiplication will be skipped~\cite{a35}. Memorization is the second subcategory of computational reduction. In \cite{a36}, memorization is achieved by storing the product of the calculated repeated patterns in a Bloom Filter (BF) unit that avoids performing the corresponding multiplications. 

Approximation units aim to perform all computations inaccurately to reduce the system's complexity. Due to the high amount of multiply-accumulate (MAC) operations required in the learning-based systems, the main focus of approximate computations is on addition and multiplication units~\cite{a12,a39,a13,appcomp1}.  
To approximate multiplication with addition, the operands are rounded to the nearest power of two using an approximation factor proposed in~\cite{a41}. 
Moreover, The logarithmic number system (LNS) method is utilized in~\cite{a46}, which translates the input values into their logarithmic equivalents, sums them, and then converts them into the linear domain.
Dynamic and static segmentation multipliers (DSM, SSM), are presented in~\cite{a43} by using m-bit input segments smaller than the input bit width in the approximate multiplication. These segments have fixed or arbitrary lengths based on the considered method. Furthermore, In~\cite{a43} a lightweight controller is proposed to combine low and high-precision approximate multipliers in the target model to adjust inaccuracy and overhead. 
In~\cite{a12} an approximate generalized matrix multiplier (AGEMM) is proposed that utilizes approximate multipliers through missing some operations during the execution.
In~\cite{a50}, the fixed point approximate matrix multiplication (FAMM) method that converts all operands to fixed point ones is proposed.  In this method, the large matrices are broken up into smaller ones, and then the results are integrated. 
In~\cite{DSI}, an adjustable approximate multiplier based on the requirements of the target system or application is proposed. This adjustment tends to achieve the appropriate balance between accuracy and resource utilization. This method splits the input into three sections: the direct part (D) is comprised of valuable bits that are carefully calculated, the average bits processed in the search section (S) are approximate, and the low-value bits that are ignored are contained in the unimportant part (I). Then it limits the computations to consider the most important parts of the number. 

Based on the reviewed studies, approximate computing-based methods efficiently regulate computation overhead and accuracy. This scheme is more flexible than algorithms and hardware compression-based methods and capable of generalizing. Extending the approximate computation to more levels of the model such as compression schemes and integrating the various approximations that provide different accuracies could improve their efficiency. 

\section{Proposed method}
\subsection{Problem Statement}
Safe driving is one of the main objectives of self-driving cars. In this context, applying high-precision algorithms to them to interact with the environment is necessary. Modern high-precision methods require powerful computation platforms that are not available in embedded systems such as the ones in self-driving cars. To this aim, in this paper designing an approximate traffic sign recognition unit called "AppSign" is aimed. The proposed unit is based on the adaptive approximations of the operations in each layer and among layers considering the final accuracy of the system. In this context, since the assumed sign detection architecture is based on CNN we focus on convolution and multiplication operations. 

Our proposed "AppSign" core unit is a CNN-based image classifier. It consists of four convolution layers performing operations on input and kernel matrices. The input matrix is the target image's feature map and the kernel is a set of weights that slide over the input to compute the output pixels. This computation is mainly based on multiplication, so approximating this operation is aimed in the proposed AppSign. Approximating at each layer and between the layers with various methods considering the effect on the final result is performed. In this context, a novel approximate multiplier called "TIRuD" is proposed to compromise the accuracy and computation overhead efficiently. Moreover, it integrates with various approximate multipliers with different precision levels to provide acceptable accuracy at the cost of low computation overhead. It should be mentioned that this unit adjusts approximation based on the criticality of the signs to avoid incorrect designs in crucial cases.

\subsection{AppSign: Our proposed Approximate Hierarchical Sign Detection Unit}
Our presented AppSign is a real-time traffic sign recognition unit for autonomous vehicles. This unit's precision should be higher due to its direct effect on the decisions of the driver-assistance system. At the same time, its computation overhead should be lower due to these systems' limited resources. In AppSign, approximate computing is employed to limit unwanted or redundant operations while keeping an acceptable accuracy. AppSign employs non-uniform approximations with different precision in all layers of the deep learning-based model to efficiently reduce the overhead of redundant or less effective operations. 

Traffic sign recognition is an image-processing process and convolutional neural networks (CNNs) are the best option to solve it due to their parameter-sharing capability. Thus, our proposed AppSign employs a four-layered CNN structure for traffic sign recognition. The majority of the operations of this structure are matrix multiplication, so we focus on their approximation. These operations are mostly based on the summing of products derived from sliding a kernel across the width and height of the input image. 

Moreover, a sensitivity analysis of the layer's operations on the final result is performed. Based on this analysis, distinct computational demands are determined for various layers of the employed CNN. The first layers of the employed CNN model have more computations due to their larger sizes leading to low sensitivity to approximate computing. This is due to high redundancy in the computations of these layers, which compensate for the effect of approximated computations on the final result. Thus, employing low-precision approximation schemes in the first layers of the model provides acceptable performance for the final recognition decision and reduces the computation overhead efficiently. 
Contrarily, the last layers of the model show more sensitivity to the approximated computations due to managing fewer operations caused by the pooling and reducing the number of parameters and computations in the network. Thus, it is required to employ high-precision computations in these final layers to achieve acceptable precision. 
As the result of these analyses, applying non-uniform approximation on the model's layer provided by combining low and high-precision approximations leads to compromised accuracy and computational overhead of the model efficiently. 

\begin{figure}
    \centering
    \includegraphics[width=0.7\linewidth]{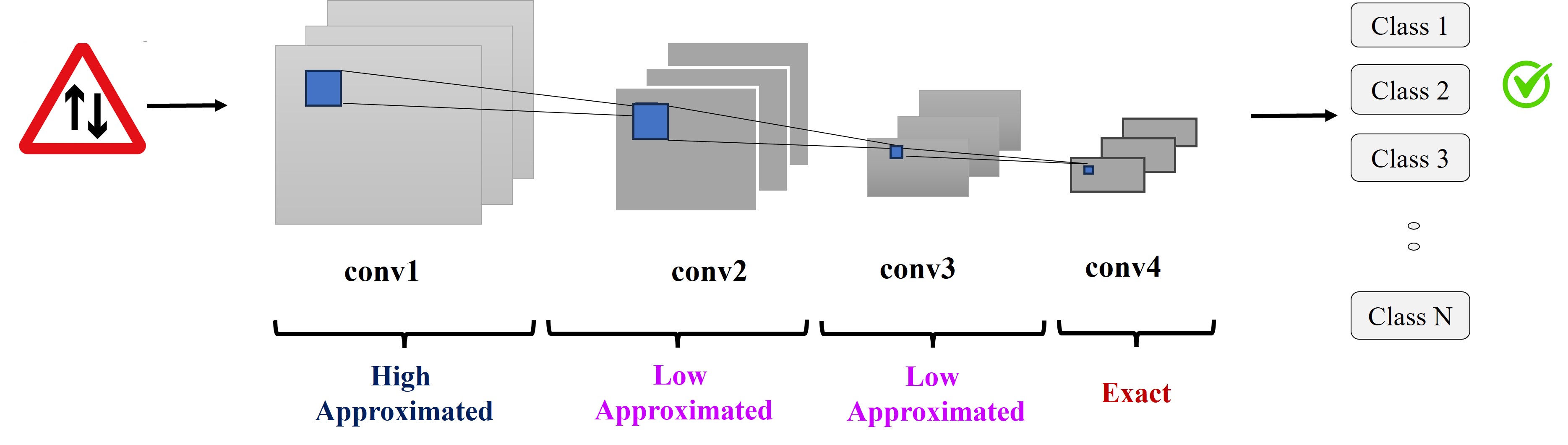}
    \caption{The architecture of the employed CNN and the approximation of the layers in AppSign regarding the computation costs}
    \label{fig:cnn}
\end{figure}

To compromise the precision and computational overhead of the employed model of AppSign, employing appropriate computation models in each layer is required. Moreover, the compatibility of the applied approximation techniques is also important. Based on the explained analysis, employing low-precision methods in the first two layers of the underlying CNN model leads to appropriate output accuracy. This estimation reduces the computation overhead dramatically due to the high number of computations in these layers but in the cost of low precision degradation. These approximation methods should be compatible as the results of the first layer are the inputs of the second one. Contrarily, the employed model of AppSign is intolerable of high approximation for the third and fourth layers, and precise computations should be employed in these layers. This non-uniform approximation agrees with the fact that the features of the final layers are more meaningful with a high abstraction level, so their approximation imprecises the final result dramatically. 
This heterogeneous combination of approximated computations keeps the overall accuracy at an acceptable level based on the target application while reducing the computational complexity. It should be noted that applying uniform approximation on all layers of the model based on previous research, biases it to high accuracy or low computation configurations which are inappropriate based on the main requirements of the target application. Fig.~\ref{fig:cnn} shows the architecture of the employed CNN and the employed approximation regarding the computation at various layers. 

\subsection{Proposed approximate multiplier (TIRuD)}
Due to the effect of multiplication in the employed traffic sign recognition unit of our proposed AppSign, we focus on this operation.
To this aim, we propose an approximate multiplier that aims to efficiently reduce complexity by removing less-effective computations.
Since generating the partial products and integrating them are the most computation-intensive parts of the multiplication, we try to limit them and keep the most effective parts. On the other hand, in the multiplication operations in CNN layers, the numbers have decimal parts due to the kernel computations. This part is eliminated in many related methods due to being undervalued, but based on our study this part reflects the main impact of the kernel on the image pixels.  
Our proposed approximation method called TIRuD (Truncated Integer Rounded Decimal) considers the decimal part of the number but with more approximation than the integer part. Moreover, its attention to different digits is regulated based on their place in the final result.

TIRuD truncates the integer part of the multiplicand first to extract its most significant digit which determines the main value of the final result. Thus, the multiplications of TIRuD in the integer part are limited to the number of digits of the multiplier. This reduces the number of partial products to the number of digits in multiplicand which is the bigger number in computation.
Thus, the integer part of the final result is approximated based on the significant digits and the low-value chunk is eliminated.  
This approximation conserves the ordering value of the result correctly which is well suited for the image processing application where the range of the numbers represents the value of each pixel.
Afterward, the decimal part of the result is computed based on rounding it to the nearest $2^n$ factor and converting the multiplication to addition. 
This approach overestimates the decimal part and regards its importance in the final result to keep the effect of accumulative kernel computation on the final result.
Alg.~\ref{alg1} shows the pseudo-code of our proposed TIRuD based on its defined two-phased scheme.    

\begin{algorithm}[!t]
\label{alg1}
\caption{Details of the proposed approximate multiplication (TIRuD)}
\small
\textbf{Inputs:} A, B   \Comment{Floating-point Operands} \\
\textbf{Output:} R \Comment{Final result} \\
\textbf{Pre-processing:} \\
$I\_A \gets [A]$ \\
$D\_A \gets A - [A]$ \\
App(D$\_$A) = $\ceil * {(log_2(D\_A)) }$
\Comment{\textit{/*Approximated D$\_$A*/}} \\
$I\_B \gets [B]$ \\
$D\_B \gets B - [B]$ \\
App(D$\_$B) = $\ceil * {(log_2(D\_B)) }$
\Comment{\textit{/*Approximated D$\_$B*/}} \\
$A \gets (A < B ? A=B)$ \Comment{\textit{/* A:multiplicand, B:multiplier*/}} \\ 
$L \gets \floor*{ (log (I\_A))}$ \\
$MSD (A) \gets I\_A[L] $ \\
\textbf{Approximate Multiplication:} \\
$D\_R  \gets 2^{(App(D\_A) + App(D\_B))}$ \\
\For {(j=0 to L-1)}{
I$\_$R$\_$tmp[j] = MSD (A) * I$\_$B [j] * $10^{L+j}$ \\
I$\_$R = $\Sigma(I\_R\_tmp[j])$} 
R = D$\_$R + I$\_$R
\end{algorithm}

Based on Alg.~\ref{alg1}, our proposed TIRuD performs separate operations on the integer and decimal parts of the operands and derives the final result by adding these separate outputs. The operation in the integer part is approximated by truncating the multiplication to the most significant digit of the multiplicand. It should be noted that the logarithm operation is performed through the shift to reduce the computation overhead.
In the decimal part, the multiplication is approximated to add by rounding up the operands to their nearest power of two. TIRuD employs a non-uniform approximation for multiplying two floating point numbers to regard the effect of each part of the operands during the computation.  

\section{Experimental result}

\subsection{Simulation setup}
\begin{figure}
    \centering
    \includegraphics[width=0.7\linewidth]{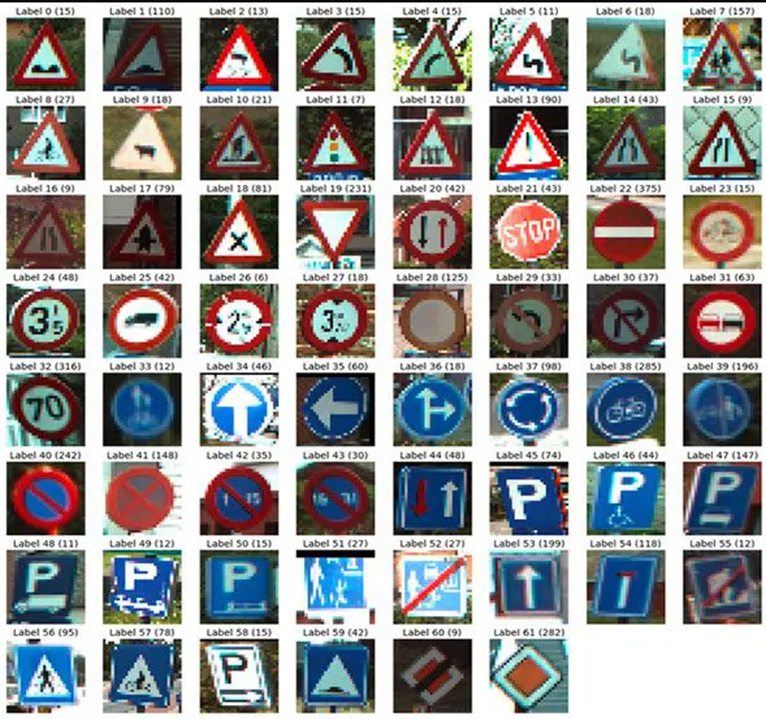}
    \caption{Samples of the traffic sign images of the employed Dataset\cite{a51}}
    \label{fig:enter-label2}
\end{figure}

To evaluate the efficiency of the proposed approach, a dataset containing more than 50,000 images of various traffic signs named GTSRB  for German Traffic Sign Recognition Benchmark is utilized \cite{a51}. This dataset has around 43 different classes of traffic signs taken in various conditions, some shown in Fig.~\ref{fig:enter-label2}.
Each class of this dataset has various images that range from 210 to 2250. The images of traffic signs are labeled with a tag that signifies their type including the speed limit, children crossing, driving on the left side, no entry, and more. 
Descriptive labels are crucial for supervised learning tasks such as image classification and recognition which is the main target of our problem.
To handle images and their associated tags, the dataset employs Python libraries like the operating system module to duplicate files and directories and the Python Imaging Library (PIL) to open and manipulate images. 
Storing each image and its tags in lists is necessary to process this dataset.
Afterward, this list should be converted to an array to feed the data to the employed model. The data input is shaped to (3, 30, 30, 39299) to represent the RGB (available color data) value, the image size in two dimensions, and the number of images. This dataset splits $20\%-80\%$ to provide various signs for learning test and train processes. 

\begin{figure}
    \centering
    \includegraphics[width=0.7\linewidth]{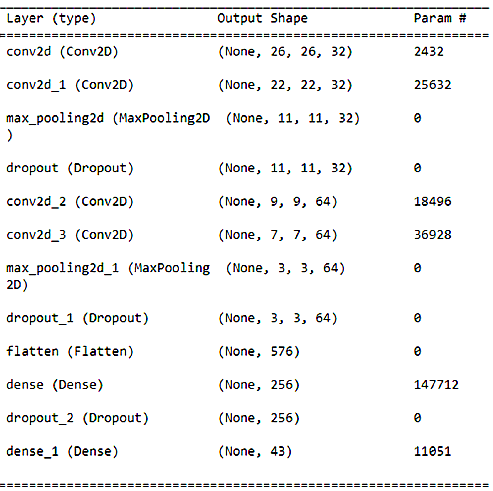}
    \caption{Summary of the employed CNN model for traffic sign recognition}
    \label{fig:enter-label3}
\end{figure}

Our considered platform for traffic sign recognition is the CNN model which classifies the images into relevant categories. This considered CNN model has four convolution layers and their detailed characteristics are shown in Fig.~\ref{fig:enter-label3}.
CNNs have the convolution layers as the main element, but their specific structures can vary. The input image is multiplied by the convolution kernel (i.e., the filter matrix). As previously stated, the convolutional layers are the primary targets for approximation within the convolutional neural network due to their high computational intensity. These layers have multipliers implemented by matrix multiplication schemes such as the GEMM (General Matrix Multiplication) function. 
This architecture achieves $95\%$ accuracy in the target traffic sign recognition problem. It should be noted that all the following experiments are implemented in Python and Google Colab. 

\subsection{Experiment}
To evaluate the effectiveness of the proposed approach several experiments are performed. These experiments are classified into four categories: assessing the efficiency of our proposed approximate multiplier (TIRuD), evaluating the proposed hierarchical AppSign in terms of the best combinations of approximation methods considering the model's accuracy, evaluating the approximated model based on joint optimization of the accuracy and computational cost that are related in a proposed metric called "AoC", and comparison to related studies.  

\subsubsection{Efficiency of Our Proposed Approximate Multiplier (TIRuD)}
As explained, our proposed TIRuD employs a non-uniform scheme to perform approximate multiplication on the integer and decimal parts of the numbers. To this end, the integer part of the multiplicand is truncated to its most significant digit. As well the decimal part is rounded to the nearest power of two to substitute the multiplication for addition.  

To evaluate the efficiency of this proposed multiplication, we have applied it to the target problem of traffic sign classification and approximate the multiplication operation of the employed CNN model. In this context, one input tensor is multiplied by the corresponding element of the other tensor in every iteration. Then the obtained result is added to the previous result to determine the convolution output. 
This process applies to all layers of the employed CNN model with approximated TIRuD and the exact multiplication to classify the traffic signs and their corresponding accuracies are shown in Table~\ref{table1}.

\begin{table}
\caption{Accuracy of the proposed TIRuD approximation and exact multiplication on various layers of the employed traffic sign recognition model}
\label{table1}
\centering
\begin{tabular}{|c|c|c|}
\hline
\textbf{Index} & 
\textbf{Accuracy Exact} &
\textbf{Accuracy TIRuD} \\
\hline
\hline
Layer1& 
$95.25\%$ &
$93.34\%$  \\
\hline
Layer2& 
$94.22\%$&
$89.88\%$  \\
\hline
Layer3& 
$93.29\%$&
$69.36\%$  \\
\hline
Layer4& 
$94.57\%$&
$21.14\%$  \\
\hline
\end{tabular}
\label{tab1}
\end{table}

As this table shows, employing approximation on the considered model reduces its accuracy according to the deepness of the layers. The shallower layers are more sensitive to approximation due to their limited data and the impossibility of the resultant compensation. Moreover, the accuracy of TIRuD is $10.67\%$ less than the exact model for the first three layers on average. This limited accuracy reduces the computational cost and execution time $54.05\%$ and $64.58\%$, respectively.
This makes our proposed TIRuD an appropriate option for real-time and resource-constrained applications such as autonomous vehicles.

\subsubsection{Efficiency of Our Proposed Multi-level AppSign}
Our proposed Appsign combines various approximate multipliers non-uniformly at different layers of the model to compromise its accuracy and computational cost. In this context, the effect of applying different approximate multipliers on the model's layer is studied in this part.    

\textbf{I) Evaluating the Efficiency of Approximate Methods on Target Model}

As mentioned, our employed model consists of four convolutional layers. In AppSign approximate computations are applied to various layers considering their effect on the final result. 
In this context, various approximation methods are applied to the model's layers separately. 
As explained in Section~\ref{related}, several methods exist to approximate the multiplication. These methods are mainly uniform and apply a specific approximation strategy to the target operation. Based on the details of these algorithms, they provide various approximation levels and computation overhead.
Approximating the fourth layer leads to very low accuracy which is not acceptable in our application. The reason behind this lies in the sensitivity of the last layer due to its limited computation which reduces the possibility of the compensation on approximate operation. Table~\ref{table2} presents the accuracy of the fourth layer of the model after applying various approximation methods. Due to these results, this layer is excluded from the approximation, and its computations are performed exactly in the following experiments. 

\begin{table}[h]
\caption{The accuracy of the considered traffic sign recognition model after applying various approximate multiplication schemes on its last layer}
\label{table2}
\centering
\begin{tabular}{|c|c|}
\hline
\textbf{Approximation Method} & 
\textbf{Accuracy} \\
\hline
\hline
Shift and xor& 
$13.61\%$\\
\hline
Shift and add& 
$6.69\%$ \\ 
\hline
LNS& 
$5.93\%$\\
\hline
FAMM& 
$6\%$\\
\hline
Quantize& 
$40.72\%$\\
\hline
Rounded& 
$5.93\%$\\
\hline
TIRuD& 
$21.14\%$\\
\hline
\end{tabular}
\label{tab1}
\end{table}

In the CNN model, the spatial input size reduces in shallower layers due to the convolution operation. This leads to a decrease in calculations in the last layer of our employed model so applying approximations to them makes the final result very imprecise ($14.28\%$ accuracy on average based on Table~\ref{table2}). Thus, we eliminate this layer from further studies in our proposed AppSign. 

Afterward, the effect of different approximate multiplication schemes on the accuracy of traffic sign recognition on the target model is studied. 
This evaluation aims to classify the approximation methods based on their accuracy level in solving our target problem. Table~\ref{table3} shows the average accuracy of various approximate multiplication methods on the first three layers of the assumed CNN model. 

\begin{table}
\caption{The average accuracy of approximate multipliers on the target CNN model}
\label{table3}
\centering
\begin{tabular}{|c|c|c|}
\hline
\textbf{Method} & 
\textbf{Average accuracy}& 
\textbf{Type} \\
\hline
\hline
FAMM& 
$85.98\%$& 
High precision\\ 
\hline
Quantize& 
$86.50\%$& 
High precision\\ 
\hline
LNS& 
$93.84\%$& 
High precision\\
\hline
Shift and add& 
$82.05\%$& 
High precision\\ 
\hline
TIRuD& 
$84.19\%$& 
High precision\\
\hline
Rounded& 
$70.70\%$& 
Low precision\\
\hline
Shift and xor& 
$73.79\%$& 
Low precision\\
\hline
\end{tabular}
\label{tab1}
\end{table}

Based on the average accuracy of this table, the considered models are classified into high and low-precision types. This categorization is performed to combine the approximate methods based on the required accuracy and computational cost of the final system more efficiently. Low-precision methods could be employed in the first layers with high computations to limit the cost and timing overhead while the redundant operations compensate for the inaccuracy. Contrary high-precision methods should be employed in shallow layers with fewer operations where the inaccuracy is closer to the final result. 
In this classification, the dividing line of the methods is the average accuracy of $80\%$ for traffic sign recognition.  

\textbf{II) Applying Approximate Computation on Multi Layers of the Target Model} 

\begin{figure}
    \centering
    \includegraphics[width=0.7\linewidth]{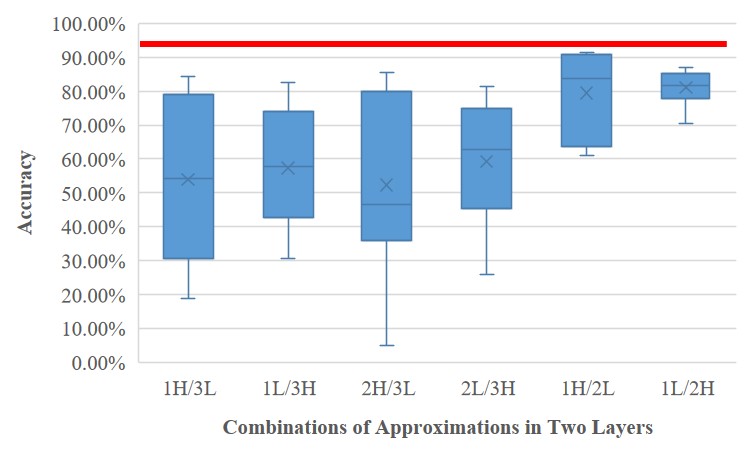}
    \caption{Statistical accuracy of employing non-uniform approximation on two convolution layers of the model. (The horizontal red line shows the accuracy of the exact method.)}
    \label{fig:enter-label4}
\end{figure}

AppSign is proposed for real-time traffic sign recognition in resource-constrained systems so its computation overhead should be minimized while the accuracy is regarded at an appropriate level.
To meet this, AppSign considers non-uniform approximation on layers of the model. Approximating the two layers of the model with low-precision and high-precision approximation methods is studied foremost.   
Different combinations of approximation methods were tested on two convolution layers of the target model, and the statistical accuracy obtained from each experiment is shown in Fig.~\ref{fig:enter-label4}.

As this figure shows, various permutations of high and low-precision approximation methods on two layers of the target model lead to different accuracies. These combinations are indicated by "H" and "L" for high and low-precision methods and a number that determines the approximated layer index. 
The statistical parameters of each combination in terms of the maximum and minimum values along with the distribution of the accuracies for various approximation methods are shown. 
Based on this figure, the best results are achieved by applying the approximation on the first two layers (the fifth and sixth bars of the figure).
Selecting the first layers of the underlying model for approximation is appropriate. Meanwhile, applying low-precision approximation on the first layer and high-precision approximation on the second layer leads to the best statistical results and is the most confident solution.  Reduction in the number of parameters and features of the shallower layers of the deep model is the main reason for this result. In these layers, limited computations are repeated for several portions, so the effect of approximations on their final result is more magnified. Thus, the final layers of the model are not good candidates for applying high approximation due to their limited and sensitive operations having certain effects on the final result.   

On condition that the acceptable accuracy of the system allows, extending the approximation to three layers of the model and reducing the computation cost is raised, 
In this context, high and low-precision approximation methods should be combined appropriately with the specified model layers. The appropriate combinations should meet the accuracy constraint and minimize computational overhead.
In this experiment, various approximation methods are applied to the first three layers of the sign recognition model, and their effects on the final output are studied. 

\begin{figure}
    \centering
    \includegraphics[width=0.7\linewidth]{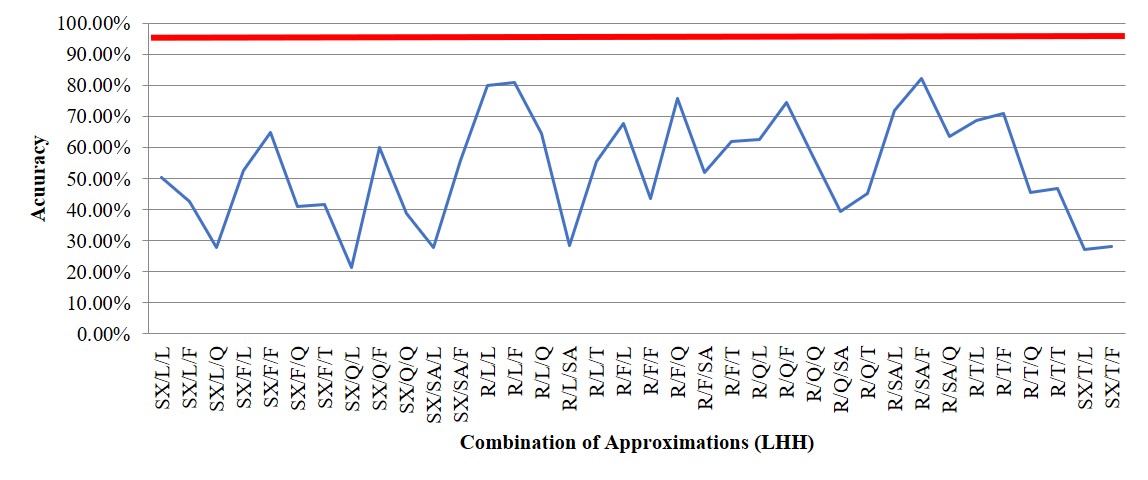}
    \caption{Accuracy of combining low, high, and high-precision approximation methods on three layers of the model (SX:Shift Xor, R: Rounding, L: LNS, F: FAMM, Q: Quantize, T: TIRuD, SA: Shift add). 
    (The horizontal red line is the accuracy of the exact method.)}
    \label{fig:enter-label6}
\end{figure}

The previous experiment shows the high sensitivity of the last layers of the target model to approximate computation. Thus no approximation is considered for the fourth layer and the third layer is bound to high-precision approximation schemes. 
Moreover, based on Fig.~\ref{fig:enter-label4} applying low precision approximation on the first layer has less accuracy degradation than the second layer. Consequently, the combinations of approximation in three layers are set to low, high, and high-precision methods, and various methods of these types are applied to the model. Fig.~\ref{fig:enter-label6} shows the result of this experiment.  

As this figure shows, the accuracy of the considered model is dependent on the types of employed approximation methods on three layers. Since their compatibility in terms of estimation scale and computations is very important. 
Based on this experiment, considering the rounding (R) approximation scheme on the first layer and LNS (L), FAMM (F), and Shift/add (SA) methods on the second and third layers leads to the best accuracy. In this context, the three prominent accuracies are for R/L/L, R/L/F, and R/SA/F combinations as $80.12\%$, $81.06\%$, and $82.35\%$ respectively. 
It should be mentioned that in this experiment, only the accuracy is considered and the computation cost of these methods is ignored. Since the main purpose of our proposed AppSign is to present a real-time traffic sign recognition unit for autonomous vehicles that are resource-constraints, the computational cost and execution time are as important as the accuracy. Accordingly, it is important to consider all these parameters simultaneously, and this point follows in the next experiment by proposing a new metric called "AoC". 

\begin{table}[h]
\caption{The average, maximum, and minimum accuracy of various combinations of approximation methods on three layers of the model}
\label{table4}
\centering
\begin{tabular}{|c|c|c|c|}
\hline
\shortstack{\textbf{Three Layers} \\ \textbf{Approximation}} 
& \shortstack{\textbf{Average} \\ \textbf{Accuracy}}
& \shortstack{\textbf{Maximum} \\ \textbf{Accuracy}}
& \shortstack{\textbf{Minimum} \\ \textbf{Accuracy}}\\
\hline
\hline
High-Low-High &
$56.55\%$ &
$79.16\%$ &
$20.14\%$ \\
\hline
Low-High-High &
$59.11\%$ &
$82.35\%$ &
$22.48\%$ \\
\hline
Low-High-Low &
$50.62\%$ &
$73.79\%$ &
$23.14\%$ \\
\hline
High-Low-Low &
$45.48\%$ &
$61.76\%$ &
$16\%$ \\
\hline
\end{tabular}
\end{table}

\begin{figure}
    \centering
    \includegraphics[width=0.7\linewidth]{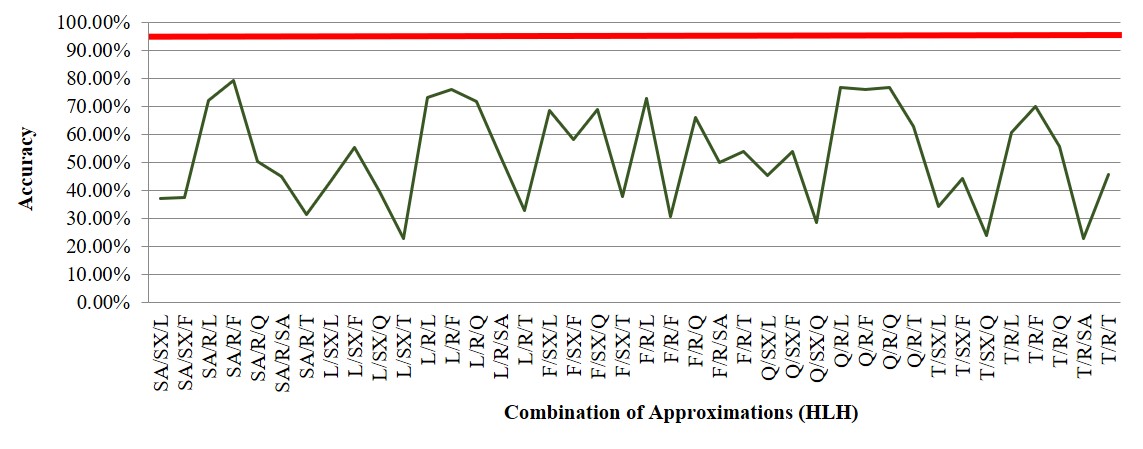}
    \caption{Accuracy of combining high, low, and high-precision approximation methods on three layers of the model(SX:Shift Xor, R: Rounding, L: LNS, F: FAMM, Q: Quantize, T: TIRuD, SA: Shift add). The horizontal red line is the accuracy of the exact method.}
    \label{fig:enter-label7}
\end{figure}

As another combination, based on the derived results of combining approximation methods for two layers, applying high-precision methods of the first layer, low-precision methods of the second layer, and high-precision methods of the third layer is considered. The accuracy of various combinations of this case is presented in Fig.~\ref{fig:enter-label7}. The parameters of this figure are set the same as Fig.~\ref{fig:enter-label6}.  

As this figure shows, the same as the previous experiment, the rounding method is the appropriate option for the low-precision layer. Moreover, LNS, Quantize, and shift/add demonstrate the best results as the high-precision approximation schemes. In this context, the three prominent accuracies are for SA/R/F, Q/R/Q, and Q/R/L combinations as $79.16\%$, $76.81\%$, and  $76.85\%$ respectively. The horizontal red line of this figure represents the accuracy of the exact method which is $94.33\%$.
Since the main target of our proposed AppSign is to optimize the accuracy and computation overhead simultaneously, the computation cost of the methods will be studied in the following experiments.   

As it has been explained, due to the previous studies and analysis, the combination of approximation methods in three layers is bound to high, low, and high-precision methods. To validate this, we have studied other combination options and summarized the results regarding the average, maximum, and minimum accuracies in Table~\ref{table4}.

\subsubsection{Compromising the Accuracy and Computation overhead}

As mentioned, employing "AppSign" in autonomous vehicles necessitates high accuracy along with low computation cost due to their real-time and resource constraint requirements. These parameters are antagonistic since one's improvement leads to the other's degradation. Thus, balancing them to an acceptable level to meet the system's precision and real-time requirements is required. 
To consider this point, we propose a new metric called "AoC" that considers the accuracy over the computation cost of the various approximation designs. This metric can evaluate the effectiveness of various combinations of approximate methods on the model's layer studied in previous experiments in terms of accuracy and computation cost, simultaneously. 
This parameter is computed as follows:  

\begin{equation}
\label{eq:poc}
\text{AoC} = \frac{\text{Accuracy}}{\text{Number of Computations}}
\end{equation}

This relation determines the accuracy percentage over computation cost. Since our target system requires maximum accuracy and minimum computation cost, the AoC should be maximized. The importance of these parameters is considered equal in AoC, but it is possible to adjust their weights based on the application requirements. 
In this context, the computation cost is considered in terms of the number of calculations. These calculations include multiplication, addition, shift, xor, and logarithm base 2 operations. Regarding simplicity, the computation costs of all operations are considered equal while we know this is imprecise due to their dependency on the processor's architecture.

\begin{table}[h]
\caption{Accuracy, computational cost, and the proposed AoC metric for various approximation methods}
\label{table5}
\centering
\begin{tabular}{|c|c|c|c|}
\hline
\shortstack{\textbf{Method} \\ \textbf{} } &
\shortstack{\textbf{Average} \\ \textbf{Accuracy}} &
\shortstack{\textbf{Average Calculations} \\ \textbf{(Thousands)}} &
\shortstack{\textbf{AoC} \\ \textbf{} } \\
\hline
\hline
Exact &
$94.33\%$ &
66.713 &
1.41397\\
\hline
LNS &
$93.84\%$ &
63.200 &
1.48481\\
\hline
Rounded &
$70.70\%$ &
32.165 &
2.19804 \\
\hline
Quantize &
$86.50\%$ &
62.780 &
1.37783 \\
\hline
FAMM &
$85.98\%$ &
50.115 &
1.71565 \\
\hline
Shift and add &
$82.05\%$ &
252.415 &
0.32506 \\
\hline
Shift and xor &
$73.79\%$ &
253.095 &
0.29155 \\
\hline
TIRuD &
$84.19\%$ &
30.650 &
2.74682 \\
\hline
\end{tabular}
\end{table}

In this context, this metric is computed for various combinations of the approximation methods to determine the best options for design. Table~\ref{table5} shows the accuracy, computation cost in terms of the number of calculations, and AoC metric for various approximation methods while employing in the first three layers of the target model, averagely. 

\begin{figure*}[h]
    \centering
    \includegraphics[width=0.8\linewidth]{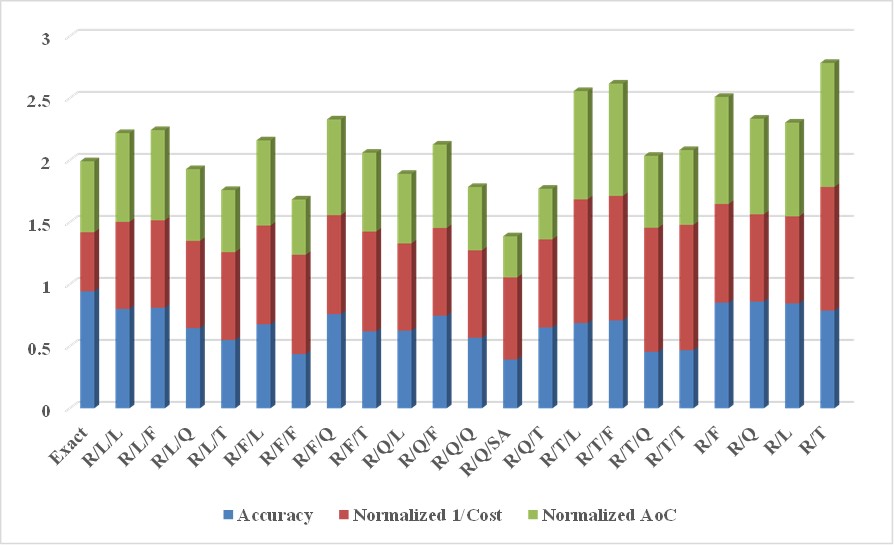}
    \caption{Comparing the combinations of approximation methods on three and two layers of the model in terms of their accuracy, normalized reversed computational cost, and normalized AoC (R: Rounding, L: LNS, F: FAMM, Q: Quantize, T: TIRuD).}
    \label{fig:enter-label8}
\end{figure*}

Based on this table, the computational cost for shift-based methods (shift/add and shift/xor) is very high. This is caused by their bitwise operations, which lead to adding several partial products. Whether these additions perform exactly or estimate with xor, the number of calculations increases dramatically. Conversely, our proposed TIRuD has the best performance considering accuracy and computation cost compared to other approximation methods. 

Regarding the presented AoC metric, the selected combinations of approximate methods in two and three layers based on the previous experiments are compared regarding their accuracy, normalized reversed computational cost, and AoC. This comparison leads to selecting the best combinations of the approximation methods that consider accuracy and computation cost simultaneously. Fig.~\ref{fig:enter-label8} shows the result of this comparison. 

In this figure, each stacked bar refers to a selected combination of approximation methods applied to three or two model's layers based on the previous experiments presented in Figs.~\ref{fig:enter-label6} and~\ref{fig:enter-label4}. The three parts of each bar are the accuracy in blue, the normalized reversed computation cost in red, and the normalized AoC in green for the selected hierarchical approximation scenarios. To facilitate the comparisons, the reverse of computation cost is considered in this figure which is ideal to be close to one. 

This figure considers 17 cases of three-layered approximation for low, high, and high-approximation combinations along with 4 cases of two-layered low, and high-estimation (the best result of Table~\ref{table4} and Fig.~\ref{fig:enter-label4}). Since all the parameters Fig.~\ref{fig:enter-label8} are set in [0,1], the proximity of each to one is ideal. 
Based on this figure, the bars of the best options that compromise the accuracy and computational cost appropriately, are taller and have higher AoC. Thus, the best three appropriate approximate models based on accuracy and computation cost in terms of AoC are R/T in the first two layers along with the R/T/F and R/T/L combinations in the first three layers of the considered sign recognition unit.
It should be noted that shift-based approximation methods are eliminated from this experiment due to their very high computational cost. 
As a result, our proposed TIRuD is an appropriate approximation method that can compromise the accuracy and computational cost in heavy deep learning-based structures employed in critical embedded systems such as autonomous vehicles.  

\subsubsection{Comparison to Related Studies}
To evaluate the efficiency of our proposed AppSign its performance should be compared to related studies. In this context, the performance in terms of accuracy, AoC, time, and the number of the models' parameters is considered. Since the proposed AppSign aims at real-time execution, these parameters are effective in evaluation. 
The related studies are categorized into approximate computing and model-based schemes. The Approximation methods of ~\cite{43} with various estimation windows and the configurable partitioned approach of~\cite{DSI} are considered in the first category. As well, two CNN-based exact methods based on attention and image enhancement models are considered~\cite{comp-r,comp-nes}.
Table~\ref{table:comp} shows the results of these comparisons.  

\begin{table}[h]
    \centering
    \caption{Comparison of the proposed AppSign with one, two, and three layers of approximation to related approximate computing and real-time CNN-based studies in terms of accuracy, response time, and AoC}
    \begin{tabular}{|c|c|c|c|}
    \hline
      Method & Accuracy & Response Time & AoC \\
       \hline
        \hline
       Exact  &  $94.33\%$ & 9.28s & 3.52 \\
        \hline
       SSM (m=4,8)~\cite{a43}  &  $78.93\%$ & 2.90s & 3.34 \\
        \hline
       DSM (m=4,8)~\cite{a43}  &  $70.19\%$ & 2.85s & 2.60 \\
        \hline
       DSI~\cite{DSI}  &  $92.42\%$ & 3.38s & 2.98 \\
        \hline
       Attention-based~\cite{comp-r}  &  \textbf{$99.71\%$} & 4.28s & 0.1 \\
        \hline
       Enhanced-based~\cite{comp-nes}  &  $99.40\%$ & 64.8s & N/A \\
        \hline
        AppSign (T) &  $93.34\%$ & 2.80s & \textbf{5.06} \\
        \hline
       AppSign (RT) &  $78.80\%$ & 0.96s & \textbf{6.22} \\
        \hline
       AppSign (RTF) &  $70.86\%$ & \textbf{0.51s} & 5.94 \\
        \hline
    \end{tabular}
    
    \label{table:comp}
\end{table}

This table compares the accuracy, execution time, and AoC of the proposed AppSign with one, two, and three layers of approximation to related studies. In AppSign, TIRuD in the first layer and combinations of the rounded, TIRuD, and FAMM schemes on two and three layers are considered. 
The approximate multiplications of~\cite{a24,DSI} apply to the first layer of the underlying model and the other layers have exact computation. Based on this table, the image enhancement-based method of~\cite{comp-nes} provides the most accuracy but in cost of high execution time and low AoC accordingly. The best execution time belongs to AppSign with three approximated layers. However, the best trade-off between the accuracy and computational cost is attained by approximating the first two layers of the model in our proposed AppSign. Moreover, this table verifies the effectiveness of our proposed AppSign and TIRuD methods in real-time action while providing appropriate accuracy. The adaptability of AppSign makes it possible to adjust accuracy and response time based on the application's requirements.

\section{Conclusion}
Compromising the accuracy and computation overhead due to the criticality and limited resources of modern applications such as autonomous vehicles is very important. Our proposed AppSign aims to resolve this trade-off by employing approximate computing in the traffic sign recognition unit of autonomous vehicles. To perform the mentioned task accurately, a CNN-based model is employed and its most frequent operation which is the multiplication is considered for approximation. 

In this context, an approximation multiplication algorithm called "TIRuD" is proposed that focuses on the integer and decimal parts of the number and reduces the computations by truncating and rounding the operands. Based on the experiments, the proposed TIRuD reduces the computations $54.05\%$ in cost of $10.67\%$ performance degradation in the first layer of the employed detection model. 

Moreover, in AppSign to solve the target problem the hierarchical approximation through combining various estimation schemes in layers of the model is considered. In this context, the approximation methods are classified based on their precision and assigned to the model's layers regarding their impact on the accuracy of the final output. The combinations of the approximate multiplication methods on two and three layers of the employed model are studied in AppSign and their corresponding accuracy and computation time are analyzed. Last, a new metric called "AoC" is proposed to compute the accuracy over computation cost of various scenarios and determine the best option based on meeting the system's trade-off. Based on this metric employing hierarchical approximation in various model layers outperforms the exact computation $20.48\%$. Considering our proposed TIRuD in the combination of approximation methods improves the performance of AoC over the exact scheme by $27.78\%$.     
It should be mentioned that the proposed approach can be extended to more operations and platforms of resource-constrained systems.
As the future trends, extending the approximation to other operations and considering partial approximation in non-critical parts of the input to improve the accuracy are proposed. 

\bibliographystyle{unsrt}

\end{document}